\documentclass[review]{elsarticle}
\usepackage{lineno,hyperref}

\usepackage{graphicx}

\journal{Journal of \LaTeX\ Templates}

\begin{document}

\begin{frontmatter}

\title{The Dynamics of Magnetic Vortices in Type II Superconductors with Pinning Sites
Studied by the Time Dependent Ginzburg-Landau Model}


\author{Mads Peter S{\o}rensen and Niels Falsig Pedersen}
\address{Department of Applied Mathematics and Computer Science,
Richard Petersens Plads, Bldg.~321, Technical University of Denmark,
DK-2800 Kongens Lyngby, Denmark} 

\ead[url]{www.compute.dtu.dk}

\cortext[mycorrespondingauthor]{Mads Peter S{\o}rensen}
\ead{mpso@dtu.dk}

\author{Magnus \"{O}gren}
\address{School of Science and Technology,
\"{O}rebro University, SE-70182 \"{O}rebro, Sweden}


\begin{abstract}
We investigate the dynamics of magnetic vortices in type II superconductors with normal state
pinning sites using the Ginzburg-Landau equations. Simulation results demonstrate hopping of vortices
between pinning sites, influenced by external magnetic fields and external currents. The system is highly
nonlinear and the vortices show complex nonlinear dynamical behaviour.
\end{abstract}

\begin{keyword}
\texttt{Ginzburg-Landau equations} \sep \texttt{type II superconductivity}
\sep \texttt{vortices} \sep \texttt{pinning sites}
\MSC[2010] 35, 37
\end{keyword}

\end{frontmatter}


\section{Introduction}
The dynamics of magnetic vortices in type II superconductors at temperatures
close to the critical temperature can be modelled by the time dependent Ginzburg-Landau equations.
The theory is based on a Schr\"{o}dinger type equation with a potential
containing a quadratic term and a quartic term in addition to a kinetic term involving the
momentum operator coupled to a magnetic field governed by the Maxwell
equations \cite{Vodolazov2004}, \cite{Lipavsky2012}, \cite{Lin1997}.
For type-II superconductors the Ginzburg-Landau equations
model the magnetic field penetration through quantized current vortices
as the externally applied magnetic field exceeds a threshold value. A
number of variants of the Ginzburg-Landau equations have been used to investigate
pattern formation in different nonlinear media, not only in superconductivity,
and hence have become a popular field of study in nonlinear science  \cite{Nielsen1973}, \cite{Scott2005}.
Our aim here is to investigate the dynamics of vortices in the presence of
normal state pinning sites in the superconductor \cite{Lin1997}, \cite{Lang2014}.
Such pinning sites can arise from atomic impurities, magnetic impurities,
lattice defects and defects in general. The Gibbs energy of the
superconductor is given by a 4'th order potential in the order parameter.
The sign of the coefficient to 2'nd order term determines the phase
transition between the normal and the superconducting state and hence
this coefficient can be used to fix the positions of inserted pinning sites.
Secondly, we shall present a model for the action of the self induced
magnetic field on vortex generation, when a net current is flowing through a
superconductor enforced by metal leads at the ends of a superconducting
strip.

\section{The time dependent Ginzburg-Landau model}

The superconducting state is described by the order parameter $\psi(\mathbf{r},t)$, where
$\mathbf{r}$ is the position in the superconducting volume denoted
$\Omega \subset R^{3}$ and $t$ is time. In the framework of the Ginzburg-Landau theory
the Gibbs energy of the superconducting state $G_s$ is given by

\begin{equation}
\label{GibbsE}
G_s = G_n -\alpha_0(\mathbf{r}) \left( 1-\frac{T}{T_c}\right)  \vert \psi \vert ^{2} + \frac{\beta}{2}  \vert \psi \vert ^{4} \; .
\end{equation}

\noindent Here $G_n$ is the Gibbs energy of the normal state, $T$ is the absolute temperature and $T_c$
is the critical temperature. The parameter $\beta$ is a constant and $\alpha_0(\mathbf{r})$ we choose
such that it depends on the space variable $\mathbf{r}$ in order to model pinning sites depleting the superconducting
state at specific positions. For $T<T_C$ positive values of $\alpha_0$ correspond to the superconducting state and
negative values model a pinning site at which the superconducting state becomes normal.

The order parameter is influenced by the magnetic field
$\mathbf{B}(\mathbf{r},t)$ given by the magnetic potential through the
relation $\mathbf{B} = \nabla \times \mathbf{A}$.
The Ginzburg-Landau parameter $\kappa$ is introduced as the ratio between the magnetic field
penetration length $\lambda$ and the coherence length  $\xi$, i.e. $\kappa = \lambda/\xi$.
We shall investigate the dynamics of flux vortices penetrating the superconductor in the
presence of pinning sites, whose positions are given by a function $f(\mathbf{r})$ taking the value
one in the superconducting regions and the value $-1$ at the position of a pinning site.
After scaling to normalized coordinates and using the zero electric potential gauge
the Ginzburg-Landau equations for the order parameter in nondimensional form reads \cite{Alstroem2011}

\begin{eqnarray}
\label{GL-1a}
\frac{\partial \psi}{\partial t} &=& -\left( \frac{i}{\kappa} \nabla + \mathbf{A} \right )^{2} \psi +
f(\mathbf{r}) \psi  - \vert \psi \vert ^{2} \psi \; ,\\
\sigma \frac{\partial \mathbf{A}}{\partial t} &=& \frac{1}{2 i \kappa} (\psi^{*} \nabla \psi - \psi \nabla \psi^{*})
- \vert \psi \vert ^{2} \mathbf{A} - \nabla \times \nabla \times \mathbf{A} \; . \label{GL-1b}
\end{eqnarray}

\noindent In order to make the Ginzburg-Landau equations dimensionless, we have scaled the space coordinates by
the magnetic field penetration depth $\lambda$ and time is scaled by $\xi^2 / D$, where $D$ is a
diffusion coefficient \cite{Gorkov68}. The magnetic field $\mathbf{A}$ is scaled by the factor $\hbar/(2e \xi)$,
where $e$ is the electron charge. The wave function $\psi$ is scaled by $\sqrt{\alpha_0 / \beta}$.
The term in $\sigma$ is the conductivity of the normal current of unpaired electrons. It is scaled by the factor
$1/(\mu_0 D \kappa^2)$, where $\mu_0$ is the magnetic permeability of the free space.
The normal current $\mathbf{J}_n$ and the super current $\mathbf{J}_s$ read

\begin{eqnarray}
\label{currents}
\mathbf{J}_n = - \sigma \frac{\partial \mathbf{A}}{\partial t} \; \; \; \mathrm{and}  \; \; \; \mathbf{J}_s =
\frac{1}{2 i \kappa} (\psi^{*} \nabla \psi - \psi \nabla \psi^{*}) - \vert \psi \vert ^{2} \mathbf{A} \; ,
\end{eqnarray}

\noindent with the total current being $\mathbf{J}=\mathbf{J}_n+\mathbf{J}_s$. The coefficient
$f(\mathbf{r})$ of $\psi$ in Eq.(\ref{GL-1a}) defines the positions of the pinning sites by changing sign from
$+1$ to $-1$ and through scaling $f$ is related to $\alpha_0$ in Eq. (\ref{GibbsE}). In solving
numerically Eqs. (\ref{GL-1a}) and (\ref{GL-1b}) we need to specify appropriate boundary conditions.
We seek to satisfy the following three boundary conditions on the boundary of the superconducting region
$\partial \Omega$

\begin{equation}
\label{BC1}
\nabla \times \mathbf{A} = \mathbf{B_{a}} \; , \; \; \; \nabla \psi \cdot \mathbf{n}= 0 \; \; \; \mathrm{and}
\; \; \;  \mathbf{A} \cdot \mathbf{n} = 0 \; .
\end{equation}

\noindent The first condition tells that the magnetic field at the surface of the superconductor equals the
applied external field $\mathbf{B}_a$. The condition $\nabla \psi \cdot \mathbf{n}= 0$
corresponds to no super current crossing the boundary. Differentiating the boundary condition
$\mathbf{A} \cdot \mathbf{n} = 0$ with respect to time shows that this condition prevents normal
conducting current to pass the boundary.

The Ginzburg-Landau equations (\ref{GL-1a}) and (\ref{GL-1b}) have been solved using the finite element
software package COMSOL Multiphysics \cite{COMSOL}, \cite{Zimmerman2006}. In order to model pinning sites
we have introduced the function
$f(\mathbf{r})$ with real values between $-1$ and $+1$, where $-1$ corresponds to a normal state
pinning site and $+1$ corresponds to regions of the superconducting state. We have chosen
$f$ to take the phenomenological form

\begin{equation}
\label{f}
f(\mathbf{r}) = \Pi_{k=1}^{N} f_k(\mathbf{r}) \; \; \; \mathrm{where} \; \;
f_k(\mathbf{r}) = \tanh((\vert \mathbf{r}-\mathbf{r}_{0k}\vert -R_{k})/w_{k}) \; .
\end{equation}

\noindent The function $f$ attains the values $-1$ around $N$ pinning sites positioned at
$\mathbf{r}_{0k}$, for $k=1,2, ... , N$. In the following we shall consider a two dimensional
superconductor where
$\mathbf{r} = (x,y)$ and $\mathbf{r}_{0k} = (x_{0k},y_{0k})$. This means that the pinning sites
are circular with radius $R_{k}$ and the transition from the superconducting state to the normal state
happens within an annulus of width $w_k$.
Assuming a two dimensional superconductor means we strictly study an infinite prism, where the currents are flowing in parallel with the $xy$-plane and the magnetic field is perpendicular to the $xy$-plane. However, the approach is valid for sufficiently
thick finite size superconductors, where the geometric edge effects are negligible. If the thickness is denoted by $t$ then $t >> \lambda$.

Other choices for modelling pinning sites are available in the literature. In particular we mention
the local reduction of the mean free electron path at pinning centres included in the
Ginzburg-Landau model by Ge et al. \cite{Ge2015}. Here the mean free path enters as a factor on the momentum
term in Eq. (\ref{GL-1a}). This approach is more based on first principles in the physical description than ours.
The above modelling strategy may also be used to investigate suppression of the order parameter.
In particular we mention the experimental work
by Haag et al. \cite{Lang2014}, where regular arrays of point defects have been inserted into a
superconductor by irradiation with He$^+$ ions. These defects acts as pinning sites.

\begin{figure}[htp]
\begin{center}
  \includegraphics[width=12cm]{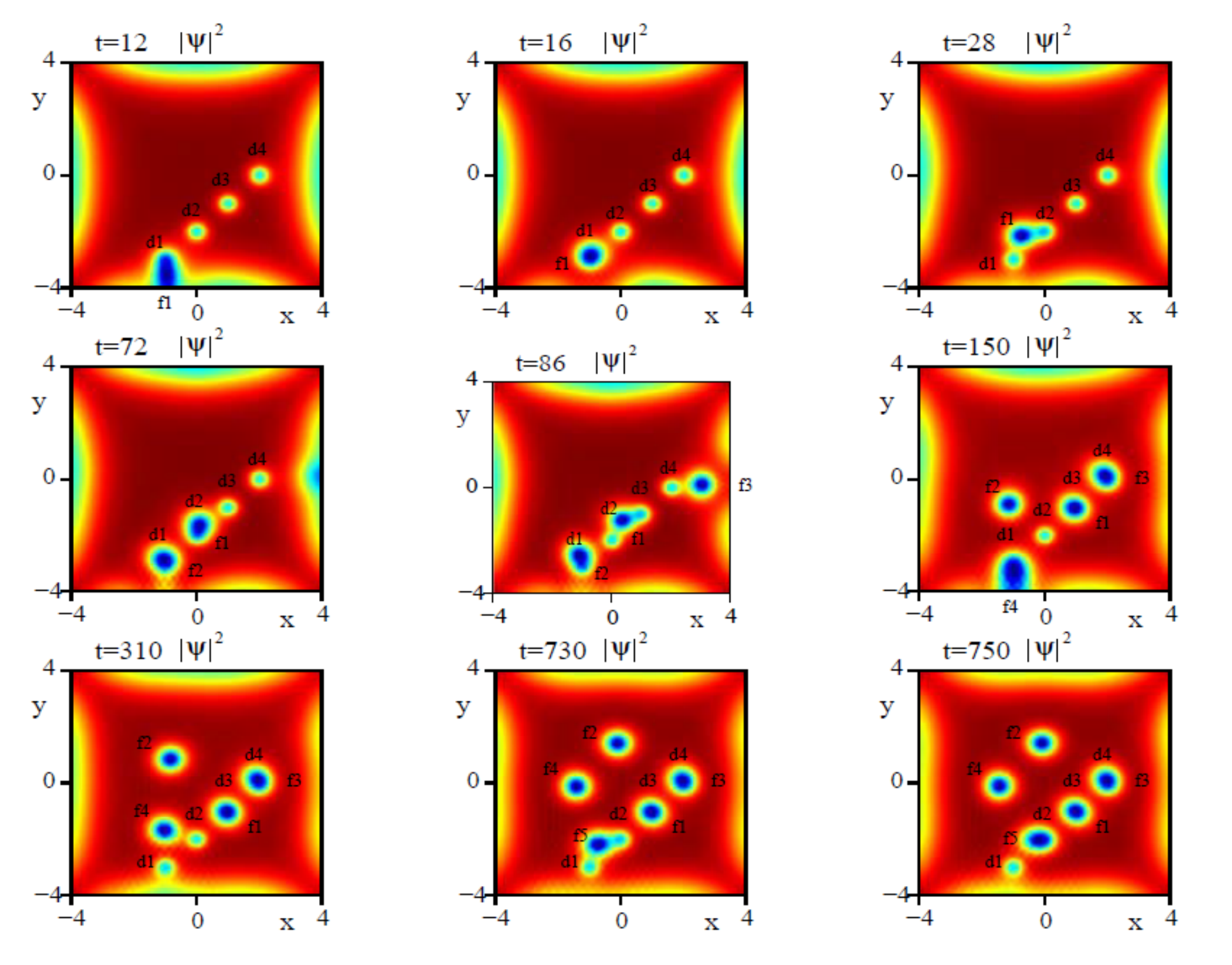}
  \caption{Numerical simulation of Eqs. (\ref{GL-1a}) and (\ref{GL-1b}) subject to the boundary conditions
  (\ref{BC1}) showing $|\psi |^2$. Dark red corresponds to $|\psi |^2=1$ and dark blue corresponds to $|\psi |^2=0$.
  The initial conditions are $\psi(\mathbf{r},0) = (1+i)/\sqrt{2}$ and $\mathbf{A}=(0,0)$.
  The parameter values are: $\kappa = 4$, $\sigma=1$, $B_a=0.73$. The positions of the pinning sites are:
  d1: $\mathbf{r}_{01}=(-1,-3)$, d2: $\mathbf{r}_{02}=(0,-2)$, d3: $\mathbf{r}_{03}=(1,-1)$ and d4: $\mathbf{r}_{04}=(2,0)$.
  }\label{fig1}
  \end{center}
\end{figure}

\paragraph{Numerical simulations} In figure \ref{fig1} we show snapshots of $|\psi |^2$ from one
simulation of the time dependent Ginzburg-Landau equations (\ref{GL-1a}) and
(\ref{GL-1b}) subject to the boundary conditions (\ref{BC1}) from time $t=0$ until $t=750$.
We have chosen the initial conditions $\psi(\mathbf{r},0) = (1+i)/\sqrt{2}$ and $\mathbf{A}=(0,0)$. The external
magnetic field $\mathbf{B}_a$ is turned on at time $t=0$. This leads to a discontinuous mismatch between the initial vanishing
magnetic field within the superconductor and the external applied magnetic field. The algorithm can handle this
without problems. Alternatively one could turn on the external magnetic field gradually giving a more smooth transition.

In the region of interest we have
inserted 4 pinning sites denoted d1, d2, d3 and d4 at the respective positions $\mathbf{r}_{01}=(-1,-3)$, $\mathbf{r}_{02}=(0,-2)$,
$\mathbf{r}_{03}=(1,-1)$ and $\mathbf{r}_{04}=(2,0)$. The pinning sites are modelled by $f$ in Eq. (\ref{f}) using $R_k=0.2$ and
$w_k=0.05$ for $k=1, 2, 3, 4$. The external applied magnetic field $B_a=0.73$ is chosen slightly smaller than the critical
magnetic field for a superconductor with no pinning sites. This value leads to very complex dynamics of fluxons entering the
superconductor resulting from mutual interactions and interactions with the pinning sites as illustrated
in Fig. \ref{fig1}. At time $t=12$ we observe a fluxon, f1, entering the superconductor, hopping from d1 to d2 influenced
by repulsive forces from the boundary and attractive forces from the pinning sites. At time $t=72$ a second
fluxon, f2, has entered the superconductor and are attached to the
pinning site d1 and eventually pushing the first fluxon f1 onto d3. As time progress the fluxon f2 deattaches d1 and moves into
the superconductor and at the same time a third fluxon, f3, enters the superconductor at the right hand side moving toward d4,
where it becomes trapped. A fourth fluxon, f4, enters at the bottom boundary close to d1 and propagates into the superconductor
and away from the pinning sites. Finally, a fifth fluxon, f5, enters close to d1 and hops from d1 to d2, where it finally
gets trapped. At $t=750$ we have obtained a stationary state with two fluxons in the bulk superconductor
and three fluxons trapped on the pinning sites d2, d3 and d4. No fluxon is attached to d1.
In short the simulation results in Fig. \ref{fig1}
illustrate the intricate nonlinear dynamical behaviour of fluxons hopping from pinning site to pinning site and at the
same time experience mutual repulsive forces and repulsive forces from the boundaries, controlled by the
external applied magnetic field.

\section{Current carrying superconducting strips}

In this section we study superconducting strips carrying currents along the strip. The current is injected
through metal contacts at two opposite boundaries of the superconductor, that is at $x=-L_x/2$ and $x=L_x/2$, respectively,
where $L_x$ is the length of the superconductor in the $x$-axis direction. 
At the side boundaries the superconducting current and the normal current are parallel to the superconductor
surface and therefore we use here the boundary conditions \cite{Vodolazov2004}, \cite{Oegren2011}

\begin{equation}
\label{BC2}
\nabla \times \mathbf{A} = \mathbf{B_{e}} = \mathbf{B}_a + \mathbf{B}_c \; , \; \; \;
\nabla \psi \cdot \mathbf{n}= 0 \; \; \; \mathrm{and}
\; \; \;  \mathbf{A} \cdot \mathbf{n} = 0 \; .
\end{equation}

\noindent In the above equations $\mathbf{B}_e$ is the total external magnetic field composed of the sum of
the applied magnetic field $\mathbf{B}_a$ and the magnetic field $\mathbf{B}_c$ induced from
the total current $\mathbf{J} = \mathbf{J}_s + \mathbf{J}_n$ in the superconducting strip.
The induced magnetic field is given by

\begin{equation}
\label{Bcurrent}
\mathbf{B}_c = \frac{1}{4 \pi} \int_{\Omega} \frac{\mathbf{J}(\mathbf{r}') \times (\mathbf{r}-\mathbf{r}')}{\vert \mathbf{r}-\mathbf{r}'\vert^3} d \Omega \; .
\end{equation}

\noindent At the metal contacts we use the metal-superconductor boundary conditions \cite{Vodolazov2004}, \cite{Oegren2011}

\begin{equation}
\label{BC3}
\nabla \times \mathbf{A} = \mathbf{B_{e}} \; , \; \; \;  \psi = 0 \; \; \; \mathrm{and}
\; \; \;  - \sigma \frac{\partial \mathbf{A}}{\partial t} \cdot \mathbf{n} = \mathbf{J}_e \cdot \mathbf{n}\; .
\end{equation}

\noindent Here $\mathbf{n}$ is the outgoing normal vector to $\partial \Omega$ and $\mathbf{J}_e$ is the external current density.
It has been shown in \"{O}gren et al. \cite{Oegren2011} that the current induced
magnetic field is well approximated by \cite{Machida1993} 

\begin{equation}
\label{BC4}
\mathbf{B}_c = \pm \frac{I}{L_y}y\mathbf{e}_z \; ,
\end{equation}

\noindent where $I$ is the total current flowing from the metal lead into and through the superconducting strip.
The width of the strip is $L_y$ and $\mathbf{e}_z$ is the unit vector in the $z$-direction
(here out of the plane). In the simulations we have used a uniform current density $\mathbf{J}_e$ given
by $\mathbf{J}_e$ = $\frac{I}{L_y}\mathbf{e}_x$, where $\mathbf{e}_x$ is the unit vector in the $x$-direction.

\begin{figure}[htp]
\begin{center}
  \includegraphics[width=5.1cm]{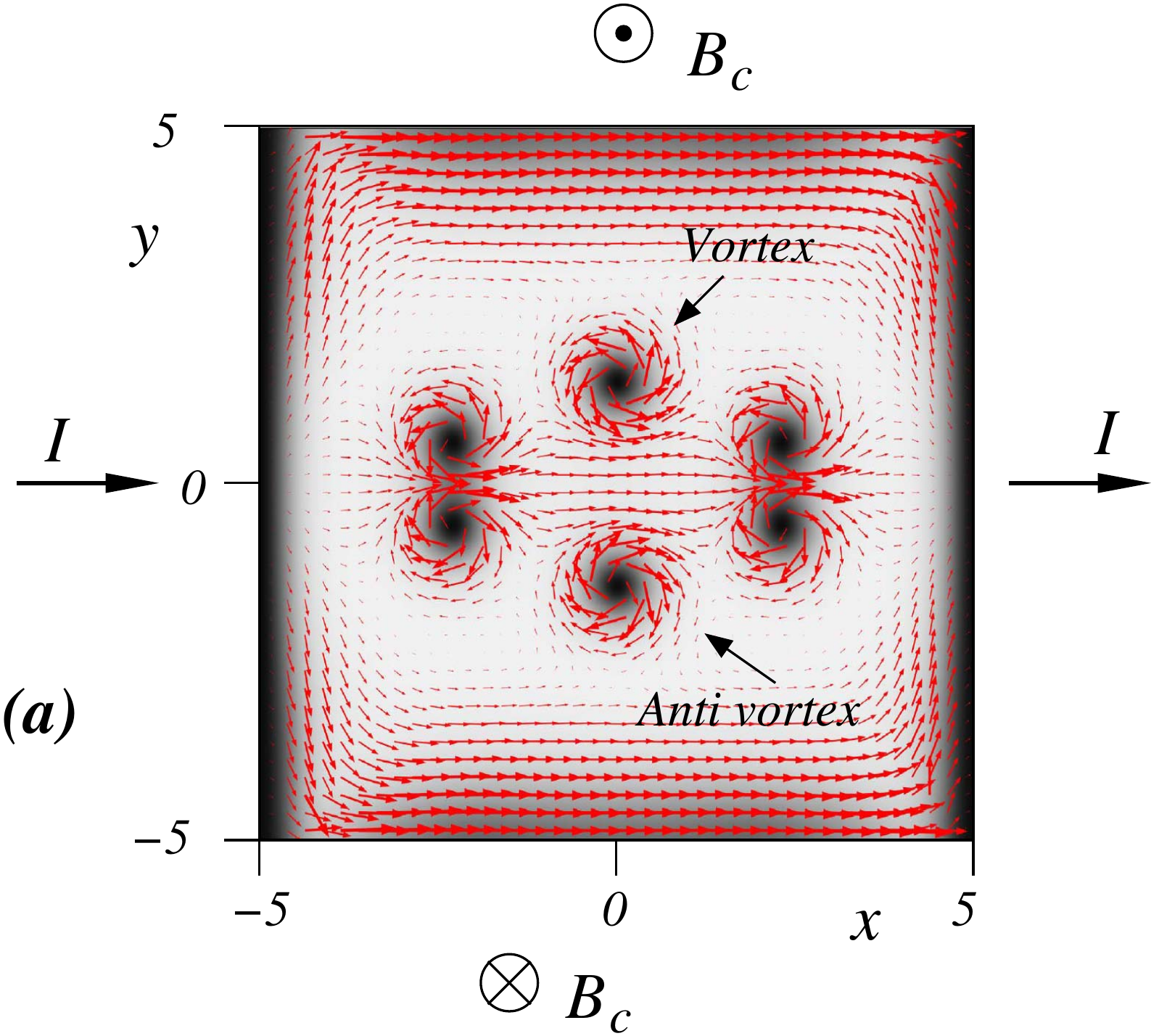} \hspace{1 cm}
  \includegraphics[width=5.1cm]{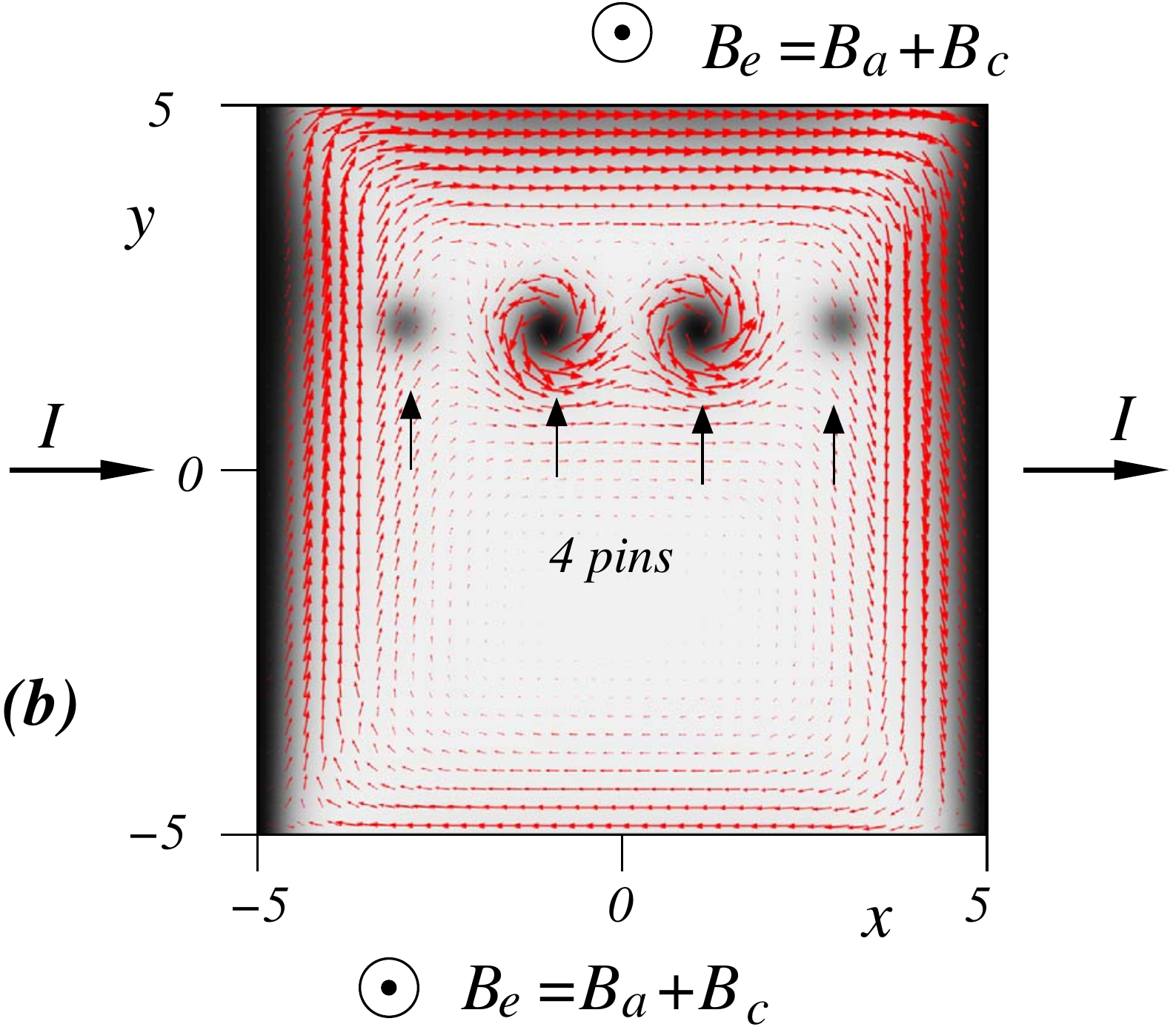}
  \caption{Numerical simulation of Eqs. (\ref{GL-1a}) and (\ref{GL-1b}) subject to the boundary conditions
  (\ref{BC2}), (\ref{BC3}) and (\ref{BC4}). The initial conditions are $\psi(\mathbf{r},0) = (1+i)/\sqrt{2}$ and
  $\mathbf{A}=(0,0)$. Parameters: $\kappa = 4$ and $\sigma=4$. (a) Simulations without pinning sites.
  Time $t=160$ using $B_a=0$ and $I=1.5$.
  (b) Simulations with pinning sites. Time $t=280$ using $B_a=0.5$ and $I=0.5$. The position of the four pinning sites are:
  $\mathbf{r}_{01}=(-3,2)$, $\mathbf{r}_{02}=(-1,2)$, $\mathbf{r}_{03}=(1,2)$ and $\mathbf{r}_{04}=(3,2)$.
  }
  \label{fig2}
  \end{center}
\end{figure}

\paragraph{Numerical simulations} Figure \ref{fig2}(a) shows how vortex anti-vortex pairs are generated at the top and
bottom boundaries, propagating into the center of a current carrying superconductor and eventually annihilate at the center.
The externally applied current, entering at the left hand side of the superconductor, generates a magnetic field at the top boundary
pointing out of the figure plane. At the bottom boundary the magnetic field points into the figure plane. No external magnetic
field $B_a$ is applied and the arrows in the figure show the strength and direction of the super current within the
superconductor. Generation of vortex anti-vortex pairs has also been studied by Milo\u{s}ovi\'{c} and
Peeters \cite{Milosevic2005} in a two dimensional superconductor structured with a lattice of magnetic dots. In this
work different complex patterns of vortex anti-vortex lattices have been found as the lattice constants of the magnetic
dots are varied.

In Fig. \ref{fig2}(b) we have inserted four pinning sites in the superconductor placed at
$\mathbf{r}_{01}=(-3,2)$, $\mathbf{r}_{02}=(-1,2)$, $\mathbf{r}_{03}=(1,2)$ and $\mathbf{r}_{04}=(3,2)$. We also apply an
external magnetic field $\mathbf{B}_a$, which adds to the induced magnetic field from the external current flowing through
the superconductor from left to right. This gives rise to an asymmetry in the magnetic field between top and bottom of the
superconductor together with an asymmetry in the current flow as is evident from Fig. \ref{fig2}(b). Using the field strength
$B_a$=0.5 and the current $I=0.5$, two fluxons enter from the top boundary and move toward the two center pinning sites, where
they get trapped. The simulations demonstrate that fluxon dynamics can be controlled by applying an external magnetic field
and by external applied currents.

\section{Conclusion}

We have modified the time dependent Ginzburg-Landau equations to model the interaction between vortices and pinning sites in a planar two dimensional superconductor. The pinning sites are modelled by multiplying the quadratic term in the Gibbs energy for the superconducting state by a suitable chosen function of the position, with a range from $-1$ at a pinning site
to $+1$ away from the pinning sites. We found that pinning sites close to the boundary of the superconductor can lower the first critical magnetic field separating the Meissner state and the flux penetration state. For magnetic fields close to the first critical
field value we found complex nonlinear dynamical behaviour of the vortices interacting with the pinning sites, mutually and with the boundaries. The vortices can hop from pinning site to pinning site influenced by repulsion from the boundaries and repulsion from other vortices, which can push pinned vortices out of a given pinning site. For interaction energies and pinning energies of
comparable magnitudes, the dynamics of vortices appears particular complex and intricate. Externally applied currents through the
superconductor also influences the dynamics of vortices in superconductors with pinning sites. Hence, the dynamics can to some extent be controlled by both external magnetic fields and external currents. We speculate that further work could
encompass derivation of particle models of fluxons in potentials modelling the pinning sites based on collective
coordinate approaches \cite{Caputo1995}.

\paragraph{Acknowledgement} We thank the EU Horizon 2020 (COST) program MP1201 Nanoscale Superconductivity: Novel Functionalities through Optimized Confinement of Condensate and Fields (NanoSC -COST) for financial support.

\section*{References}


\end{document}